\documentclass[epj]{svjour}

\usepackage{amsmath,amssymb,bm}
\usepackage{epsfig}
\usepackage{graphicx}
\usepackage{amsfonts}
\usepackage{epstopdf}
\usepackage[usenames,dvipsnames]{color} 
\usepackage{float}

\newcommand{\eq}{\begin{equation}}
\newcommand{\eeq}{\end{equation}}
\newcommand{\bd}[1]{ \mbox{\boldmath $#1$}  }

\def\ii{\'i}
\newcommand{\beqa}{\begin{eqnarray}}
\newcommand{\eeqa}{\end{eqnarray}}

\begin{document}

\title{
$^{16}$O within the Semimicroscopic Algebraic Cluster Model and the importance of the Pauli Exclusion Principle
}

\author{
P. O. Hess\inst{1,2}, J. R. M.  Berriel-Aguayo\inst{1} and  L. J. Ch\'avez-Nuñez\inst{1}
	}

\institute{Instituto de Ciencias Nucleares, Universidad Nacional Aut\'onoma de M\'exico, 
Ciudad Universitaria, Circuito Exterior S/N, A.P. 70-543, 04510
M\'exico D.F. M\'exico
\email{hess@nucleares.unam.mx}
\and Frankfurt Institute for Advanced Studies, Johann Wolfgang Goethe
 Universit\"at, Ruth-Moufang-Str.1, 60438 Frankfurt am Main, Germany  
}

\date{Received: date / Revised version: date}
%
\abstract{
The {\it Semimicroscopic Algebraic Cluster Model} (SACM) is applied to $^{16}$O, assumed to consist of
a system of four $\alpha$-clusters. 
For the 4-$\alpha$ cluster system a microscopic model space  is 
constructed,
which observes
the {\it Pauli-Exclusion-Principle} (PEP) and is symmetric under permutation of the 4$\alpha$-particles. 
A phenomenological
Hamiltonian is used, justifying the name {\it Semi} in the SACM.
The spectrum and transition values are determined.
One of the main objectives is to test the importance of the 
{\it Pauli Exclusion Principle} (PEP), comparing the results with
the {\it Algebraic Cluster Model} (ACM), which does not include the PEP, and claims that
the $^{16}$O shows evidence of a tetrahedral structure, which can be explained easily by symmetry arguments.
We show that PEP is very important and cannot be neglected, otherwise it leads to a wrong
interpretation of the band structure and to too many states at low energy.
} 

\maketitle
\section{Introduction}
\label{intro}
$^{16}$O nucleus is a test case for many structural investigations, as for example microscopic
full scale calculations \cite{DrO16,dytrych}, study of $\alpha$-cluster condensation \cite{schuck2013,schuck2017,funaki2018}.
Reviews are published in \cite{schuck2017,THSR,freer2017,review-clusters} and still open problems are discussed
in \cite{funaki2010,schuck2018}. All of them use the PEP as a fundamental principle.
In \cite{bijker2016,bijker2017} the {\it Algebraic Cluster Model} (ACM)  is applied to $^{16}$O, not taking into account
the PEP. The claim is that experimental information supports a tetrahedral structure of $^{16}$O in its
ground and excited states. The spectrum and several electromagnetic transition values are calculated
and the agreement to experiment seems acceptable.

The similar claim was made in \cite{bijker2014,annphys} for the $^{12}$C nucleus having a triangular structure.
Experimental data seems to be well reproduced. In \cite{bijker2014} the additional measured $5^-$ state
was argued as an evidence for the triangular structure.
 
In \cite{hess2018} the $^{12}$C nucleus was reexamined, within the SACM \cite{sacm1,sacm2}, with the same objective as here, namely to investigate
the importance of the PEP and changes in structure when it is taken into account. The main finding was that the spectrum and 
transition values can be equally reproduced as in \cite{annphys}, in fact there are several procedures which can
(see e.g. \cite{THSR}). It was shown that PEP is important: Not taking into account the PEP leads to too many
states at low energy, the association into bands as done in \cite{annphys} is wrong, not supported by the
shell model, and the geometrical interpretation is trivial. The measured $5^-$ state \cite{bijker2014} is not an evidence
for the triangular structure, and in addition, a further $5^-$ state predicted by the ACM at low energy is not supported
by the shell model.  Using the symmetry character of the $\alpha$ particles under
permutation, in the ground state the configuration has to be a triangle, however not necessarily for the excited
states, as for the Hoyle state \cite{hoyle}. 

In this contribution we investigate if the claims in \cite{bijker2016,bijker2017} are justified. We will use again the
{\it Semimicroscopic Algebraic Cluster Model} \cite{sacm1,sacm2}, whose model space observes the PEP but whose Hamiltonian is
phenomenological. The advantage of this model is its easy application, allowing large scale calculations
(complete spectrum and transition values), and 
its similarity to the model used in \cite{bijker2016,bijker2017}. We will concentrate on the energies, $B(EL)$ ($L=2,3$) values 
for making a point, namely that the ACM is inconsistent in not taking into account the PEP. The calculation of
more experimental data does not improve the situation. 

In section \ref{sec2} the model space of $^{16}$O is constructed, in section \ref{sec3} the Hamiltonian presented 
and results are shown. In section \ref{sec4} Conclusions are drawn. 

\section{The model space for $^{16}$O}
\label{sec2}

In this section the SACM model space is constructed for a completely symmetric 4-$\alpha$ particle system,
satisfying the PEP. The procedure can be reduced to a pure counting, avoiding cumbersome calculations of kernels.
We construct {\it ket}-states, classified by their quantum numbers.

In a system of four identical particles, 3 Jacobi coordinate vectors are introduced, leading to invariance in spatial 
translations:

\beqa
{\bd \lambda}_1 & = & \frac{1}{\sqrt{2}}\left( {\bd x}_2 - {\bd x}_1\right)
\nonumber \\
{\bd \lambda}_2 & = & \frac{1}{\sqrt{6}}\left( 2{\bd x}_3 - {\bd x}_1- {\bd x}_2\right)
\nonumber \\
{\bd \lambda}_3 & = & \frac{1}{\sqrt{12}}\left( 3{\bd x}_4 - {\bd x}_1- {\bd x}_2- {\bd x}_3\right)
~~~,
\label{jacobi}
\eeqa
where ${\bd x}_k$ denotes the position vector of the $k$'th particle. 
 The first one, 
${\bd \lambda}_1$, is the relative vector
between the first two particles, the second one, ${\bd \lambda}_2$,  is a vector from the third particle to the center of mass
of the first two and the last Jacobi vector, ${\bd \lambda}_3$, connects the fourth particle to the center of mass of the first three.
For each of the Jacobi vectors we can associate an harmonic oscillator, with quantum numbers
$(N_{\lambda_k},l_{\lambda_k},m_{\lambda_k})$, $N_{\lambda_k}$ being the number of oscillation quanta, 
$l_{\lambda_k}$ is the angular momentum and $m_{\lambda_k}$ is its projection.
For each of the sub-systems the Wildermuth condition \cite{wildermuth} has to be satisfied. The Wildermuth condition
gives the minimal number of relative oscillation quanta, necessary for observing the PEP. In the case of $^{16}$O,  
for the two-particle
sub-system the minimal number of quanta is 4, the same for the second and the third oscillator.

The definition (\ref{jacobi}) does not consider that the four particle state has to be symmetric under permutation, thus
one can use also any other combination for the Jacobi coordinates, permuting the indices. In order to do so, in what follows we indicate
the construction of completely symmetric states, which however does not yet include the Wildermuth condition. This will be done
in a subsequent  step.

In \cite{kramer-II} the states for 4 identical particles was constructed for an arbitrary permutationl symmetry, while
in \cite{mosh-book} the basic method is explained, though only angular momentum zero states were considered. Here,
we outline the basic steps, for a complete symmetric system:

New coordinates where defined
in \cite{kramer-II}, i.e.,

\beqa
{\ddot{\bd x}}^1 & = & \left( {\bd x}^1 +{\bd x}^4 -{\bd x}^2 -{\bd x}^3 \right)
\nonumber \\
{\ddot{\bd x}}^2 & = & \left( {\bd x}^2 +{\bd x}^4 -{\bd x}^1 -{\bd x}^3 \right)
\nonumber \\
{\ddot{\bd x}}^3 & = & \left( {\bd x}^3 +{\bd x}^4 -{\bd x}^1 -{\bd x}^2 \right)
~~~,
\label{coord}
\eeqa
which have more useful properties under permutation.

The $S_4$ permutation group can be written as a semi-direct product $D_2 \wedge S_3$, where the $D_2$ is composed
of the permutations

\beqa
&
e, ~~ d_1=(1,4)(2,3), ~~ d_2 =(2,4)(1,3), &
\nonumber \\
&
d_3=(3,4)(1,2)
~~~,
&
\label{d2}
\eeqa
where $(i,j)$ denotes the permutation of $i$ with $j$ and $e$ is the unit element. The $S_3$ is the
permutation group of the first three coordinates ${\bd x}^1$, ${\bd x}^2$ and ${\bd x}^3$.  A general permutation can be written
as the product of an element of $D_2$ with $S_3$.

Restricting to a complete symmetric state (as it is the case fora 4 -$\alpha$ system), the state will be multiplied by the factor
\cite{kramer-II}

\beqa
\left(1+(-1)^{{\ddot l}_1+{\ddot l}_2}+(-1)^{{\ddot l}_1+{\ddot l}_3}+(-1)^{{\ddot l}_2+{\ddot l}_3}
\right)
~~~,
\eeqa
where ${\ddot l}_k$ is the orbital angular momentum of the oscillator basis state. This factor is only different from zero, if
all ${\ddot l}_k$ are even or odd. This also implies that the oscillation quantum numbers ${\ddot N}_k$ (related to the coordinate
${\ddot{\bf x}}^k$) have to be either all even or all odd.

The general expression for the basis state is

\beqa
\mid \left[( {\ddot N}_1 {\ddot l}_1, {\ddot N}_2 {\ddot l}_2)l_{12}; {\ddot N}_3{\ddot l}_3\right]\rho  (\lambda \mu) \kappa L M\rangle
~~~. 
\label{state}
\eeqa
A state with ${\ddot N}_k$ oscillation quanta transforms as $({\ddot N}_k,0)$ with respect to $SU(3)$, $(\lambda ,\mu )$ is a
general $SU(3)$ irreducible representation (irrep), $\rho$ and $\kappa$ is a multiplicity indices, $l_{12}$ the intermediate angular momentum to which the first two oscillator 
functions are coupled, $L$ is the total angular momentum and $M$ its projection. 
In order to avoid a double counting, an order has to be followed.
One possibility is to use

\beqa
{\ddot N}_1 \ge {\ddot N}_2 \ge {\ddot N}_3
~~~.
\label{order}
\eeqa
Thus the preliminary list of $SU(3)$ irreps for a 4-particle system is obtained by multiplying 
$({\ddot N}_1,0)\otimes ({\ddot N}_2,0) \otimes ({\ddot N}_3,0)$, with the order given in (\ref{order}), 
${\ddot N}_1+{\ddot N}_2+{\ddot N}_3=N$, $N$ being the
number of oscillation quanta in the system considered and taking into account that when $N$ is even (odd), 
also all $ {\ddot N}_k$ have to
be even (odd) \cite{kramer-II}. Another restriction enters when $({\ddot N}_1,0)$ is coupled with $({\ddot N}_2,0)$ to
certain $SU(3)$ irreps $(\lambda , \mu )$:
For ${\ddot N}_1 = {\ddot N}_2$, the $SU(3)$ Clebsch-Gordan coefficient involved requires by symmetry that
$\lambda + \mu =$even.

In such a manner, for each number of total oscillation quanta, $N$, a list of $SU(3)$ irreps $(\lambda \mu )$ and their
multiplicity $\rho$ of appearance is obtained.

This list still contains too  many irreps. 
One has to subtract from it all irreps which violate the {\it Wildermuth
condition} \cite{wildermuth} and we determine it in the following:

First, we have $N_{\lambda_1}+N_{\lambda_2}+N_{\lambda_3}$ = $N$, with 
$N_{\lambda_1}\le N_{\lambda_2}\le N_{\lambda_3}$. The coupling of $(N_{\lambda_1},0)$ with  $(N_{\lambda_2},0)$,
for $N_{\lambda_1}=N_{\lambda_2}$,
underlies the same symmetry restriction, as discussed above.
Each combination of the $N_{\lambda_k}$, where at least
one of the $N_{\lambda_k}$ is lower than allowed by the Wildermuth condition, provides a condition for which the state has
to vanish. Thus, if a given $SU(3)$ irrep, $(\lambda , \mu )$ appear with a multiplicity $\rho$ in the list of completely
symmetric states, each single condition of a not allowed combination of the $N_{\lambda_k}$ reduces this multiplicity by one. 
There may be more conditions for the same $(\lambda , \mu )$. If there are $\delta$ of those conditions,
the final multiplicity of the $(\lambda , \mu )$ irrep is $(\rho - \delta)$. If this number is lower or equal to zero, the 
irrep in question is skipped completely from the list.

The procedure just explained provides a list of $SU(3)$ irreps for a given excitation number $\Delta n_\pi$, 
having already taken into account the Wildermuth condition as a minimal one for observing the PEP. However,
this list still contains irreps which are not allowed by the Pauli exclusion principle. For example, for 
$\Delta n_\pi = 0$, one obtains by the above explained procedure 
the (2,2), (1,1) and (0,0) irreps, from which only the last one is permitted. The final step is to 
apply the method of the SACM and look at the {\it overlap} of the list with shell
model states.  We will only consider states
of the so-called {\it super-multiplets} \cite{supermultiplets}. A super-multiplet is
characterized by the "most antisymmetric" irreducible representation (irrep) of the spin-isospin $U(4)$ group, i.e. for a
system of $4n$ particles, it is of the form $[n^4]$, for $(4n+1)$ it is $[5,4^n]$, etc., they
correspond for $4n$ particles to spin-isospin zero states.
To be more specific, considering
a particular shell number $\eta$ with an occupation number $A_\eta$. The relevant group chain for classification is

\beqa
U\left(4*\frac{1}{2}\left(\eta +1\right)\left(\eta+2\right)\right) & \supset 
~ U\left(\frac{1}{2}\left(\eta +1\right)\left(\eta+2\right)\right) & \otimes U_{ST}\left( 4 \right) 
\nonumber \\
\left[ 1^{A_\eta}\right]~~~~~~~~~~~ &  ~~~\left[{\widetilde h}\right] & ~~\left[ h \right]
~~~,
 \nonumber \\
\label{eq1}
\eeqa
where $[h]$ is the Young diagram of the $U(4)$ group and $[{\widetilde h}]$ the conjugate diagram, where rows and columns are
interchanged. The $S$ and $T$ refer to spin and isospin quantum number respectively.
The $SU(3)$ group is contained in the orbital group $U\left(\frac{1}{2}\left(\eta +1\right)\left(\eta+2\right)\right)$, i.e., in
order to get the shell model content one has to apply a reduction. Programs for that are available \cite{aka,bahri}.
In a standard manner the center of mass spurious motion is subtracted.

With that, we are lead to the list of $SU(3)$ irreps as given in Table \ref{tab1}, for up to 4 excited quanta.

\begin{center}
\begin{table}[h!]
\centering
\begin{tabular}{|c|c|}
\hline\hline
$n\hbar\omega$ & $\left(\lambda , \mu\right)$ \\
\hline
0 & (0,0) \\
1 & (2,1)  \\
2 & (2,0), (3,1), (4,2)  \\
3 & (3,0), (4,1), (1,4), (5,2), (6,3)  \\
4 & (0,2), (1,3), (4,0)$^2$, (3,2)$^2$, (2,4), \\
   & (5,1)$^3$, (4,3), (3,5), (6,2)$^2$, (7,3) \\
\hline
 \end{tabular}
\caption{
Model space of the 4-$\alpha$-particle system,
for up to $4\hbar\omega$ excitations and only including states with the
"most antisymmetric" irrep in $U_{ST}(4)$. 
Multiplicities larger than 1 are indicated by an upper index. 
} 
\vspace{0.2cm}
\label{tab1}
\end{table}
\end{center}

Already from Table \ref{tab1} the grouping of states into bands can be deduced within a pure $SU(3)$ symmetry
(see also \cite{elliott}):
Each band is classified by their $SU(3)$ irrep $(\lambda , \mu )$. As shown in \cite{rowe,castanos1988}, the quadrupole deformation
associated to each irrep is given by

\beqa
\beta & = &
\left[
\left( \frac{4\pi}{2N_s^2}\right)\left( \lambda^2 + \lambda \mu +\mu^2 +3\lambda + 3\mu \right)
\right]^{\frac{1}{2}} 
~~~.
\label{def}
\eeqa
The deformations for the $(0,0)$ is 0, for $(3,1)$ it is 0.36 and for $(4,2)$ it is 0.49, i.e.,
each irrep has a completely different deformation and, thus, defining a different band.

\section{The Hamiltonian and }
\label{sec3}

\subsection{The Hamiltonian}

As a Hamiltonian we propose a combination of a pure $SU(3)$-part
and a symmetry breaking term, which is a generator of $SO(4)$. 
There are more general ones, but we want to keep it simple enough with the numer of parameters as in \cite{bijker2017}.
The choice does not exclude more general Hamiltonians.
As in \cite{bijker2016} the total number of quanta is 
$N=n_{\lambda_1} + n_{\lambda_2}+ n_{\lambda_3} + n_\sigma$ = $n_\pi+n_\sigma$, where the $\sigma$-bosons are
introduced as a trick to obtain a cut-off and $n_\pi$ refers here to the total
number of excited quanta in the relative motion. The ${\bd \pi}_m^\dagger$-boson operators ($m=0, \pm 1$)
are the creation and the ${\bd \pi}^m$ operators are
the annihilation operators for the total relative motion. The algebra is $U(3)$ and their generators 
$C_m^{m^\prime}$ are the sum of generators of the three harmonic oscillators related to the Jacobi coordinates.

For $^{16}$O, the model Hamiltonian proposed is

\beqa
{\bd H} & = & \hbar\omega {\bd n}_\pi
-\chi \left(1 -\xi \Delta {\bd n}_\pi \right)  {\bd {\cal C}}_2(\lambda , \mu ) 
+ t_1{\bd {\cal C}}_3(\lambda , \mu )
\nonumber \\
&& + t_2({\bd {\cal C}}_2(\lambda , \mu ))^2 
+ \left(a+ a_{Lnp} \Delta {\bd n}_\pi \right) {\bd L}^2 +b{\bd K}^2
\nonumber \\
&& +b_1 \left[\left({\bd\sigma}^\dagger \right)^2 
- \left( {\bd \pi}^\dagger \cdot {\bd \pi}^\dagger \right) \right]
\cdot \left[h.c.\right]
~~~.
\label{eq-6}
\eeqa
The  first term is just the harmonic oscillator field and the $\hbar\omega$ 
is fixed 
via $45 \times A^{-1/3} - 25 \times A^{-2/3}$ \cite{hw}
(units in MeV), 
where its value is 
13.92~MeV
for $^{16}$O. The second term is proportional to the quadrupole-quadrupole interaction.
The intensity of this interaction is
modified with increasing shell model excitation. The third term is the third-order Casimir operator, which allows to
distinguish between $(\lambda , \mu )$ and $(\mu , \lambda )$. In the second line, the first term is the square of the
second-order Casimir operator of $SU(3)$. The second term is the ${\bd L}^2$ term.
The factor describes changes in the moment of inertia as a function of the shell excitation. The
third term distinguishes between angular momentum states in the same $SU(3)$ irrep. Finally, the last line contains the
$SU(3)$-$SO(4)$ mixing interaction.

Up to the second line, the Hamiltonian is within the $SU(3)$ limit
and permits analytic results, substituting the operators by their 
corresponding eigenvalues.
The term in the last line mixes $SU(3)$ irreps, it is
a generator of a $SO(4)$ group. 
The pure $SU(3)$ part has 7 free parameters and including the $b_1$, corresponding to 
$SO(4)$ mixture, there are in total 8 parameters. This is the same number as  in the $^{12}$C case \cite{hess2018}.
As we will see in the applications, the fits require $a=0$, so in fact there is one parameter 
less.

The physical quadrupole operator \cite{octavio,sympl} is given by

\beqa
{\bd Q}^{phys}_{2m} & = & {\bd Q}^a_{2m} + \frac{\sqrt{6}}{2}
\left( {\bd B}^\dagger_{2m} + {\bd B}_{2m} \right)
\nonumber \\
{\bd B}^\dagger_{2m} & = & \left({\bd \pi}^\dagger \cdot {\bd \pi}^\dagger
\right) 
~,~
{\bd B}_{2m} ~=~ \left({\bd \pi} \cdot {\bd \pi} \right)
\nonumber \\
{\bd Q}^a_{2m} & = & \sqrt{6}\left[ {\bd \pi}^\dagger 
\otimes {\bd \pi}\right]^2_m
~~~.
\label{Qphys}
\eeqa
The ${\bd B}^\dagger_{2m}$ operator transforms as a (2,0) $SU(3)$ irrep,
while ${\bd B}_{2m}$ as its conjugate.

\subsection{Remarks on a geometrical mapping}

In \cite{hess2018} a geometrical mapping of the Hamiltonian for the $^{12}$C nucleus was presented.
The main steps \cite{geom} are to define a trial state $\mid \alpha\rangle$ and the classical potential
as the expectation value of the Hamiltonian with respect to this trial state, i.e., \cite{geom}

\beqa
V(\alpha ) & = & \langle \alpha \mid {\bd H} \mid \alpha \rangle
~~~.
\label{pot}
\eeqa 

In the case of $^{12}$C it was found that in the ground state by default a triangular structure of the three-$\alpha$
particle system has to be realized. The argument relies on the fact that the $\alpha$-particles are indistinguishable
and the distance of the first to the second particle has thus to be equal as the distance of the first to the third and the
second to the third particle. One has to keep in mind that the notion of "first", "second" and "third" does not make sense
due to the indistinguishably of the $\alpha$-particles, it is only a {\it classical} picture.

The same can be applied to the $^{16}$O nucleus as a system of four $\alpha$-particles: The distance of any two 
$\alpha$-particle has to be the same. The only geometrical figure, which satisfies it, is the tetrahedral structure.

The situation changes when excited states are considered, which allows the mixing of different geometrical
configurations and as a result the tetrahedral structure is lost in general.

The important point is that even when the density profiles show a tetrahedral structure (see for example
\cite{kanada}) and a {\it classical} description of this symmetry leads to a classification according to the
$T_4$ dynamical group, the projection of these {\it classical} wave functions onto antisymmetrized states leads to
destroy the association into bands according to $T_4$. For example, antisymmetrization of the ground state leads to a spherical
nucleus and the band associated to it consists only of one and one state.

To resume, the methods used in atomic molecular physics cannot be translated without cause to nuclear physics, where the
PEP plays an essential role. In atomic molecular physics there is no overlap of the wave function between each atom 
nucleus but in  nuclear physics there is.

\subsection{Results}

\begin{figure}
\begin{center}
\rotatebox{270}{\resizebox{120pt}{120pt}{\includegraphics[width=0.23\textwidth]{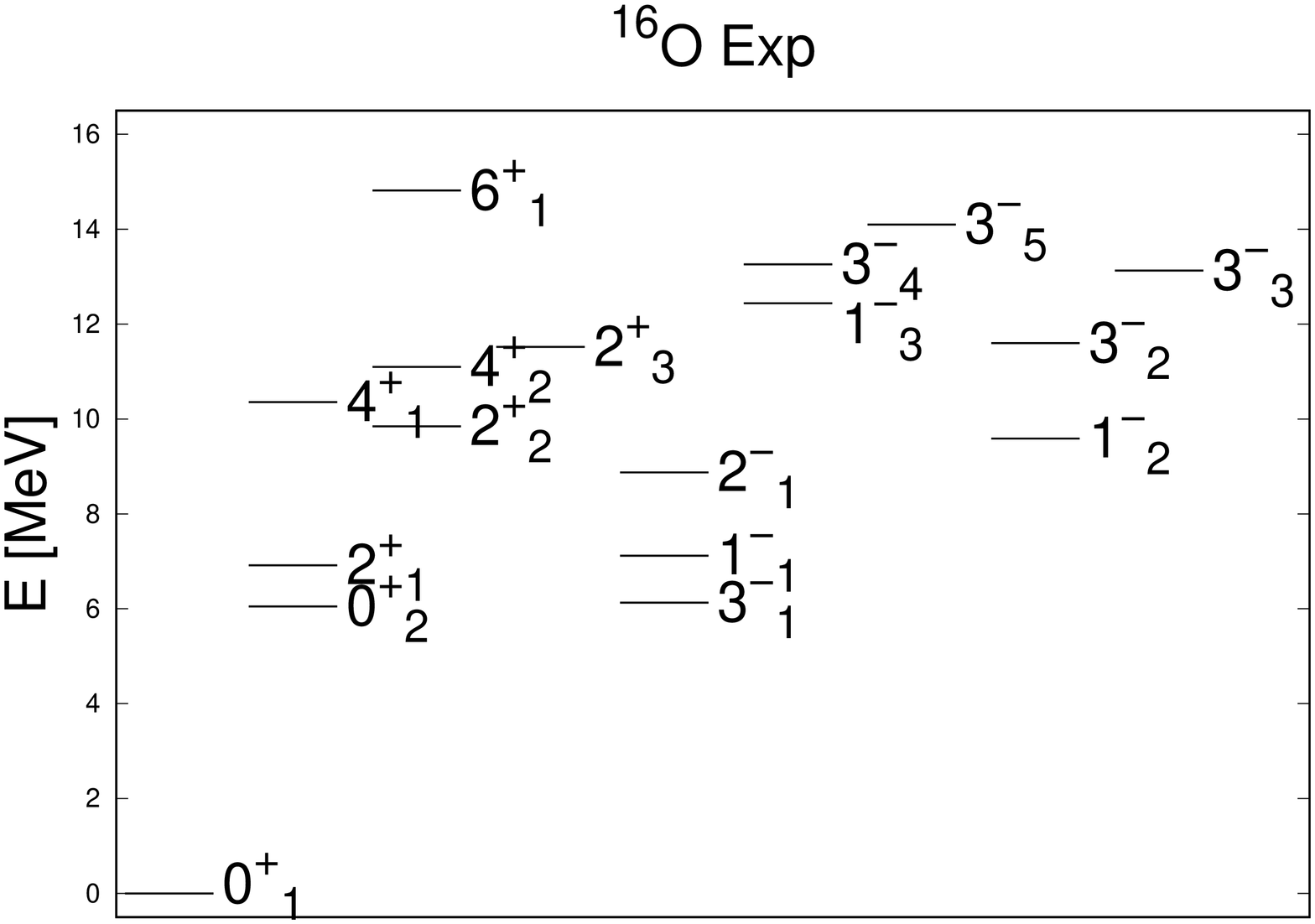}}} 
\rotatebox{270}{\resizebox{120pt}{120pt}{\includegraphics[width=0.23\textwidth]{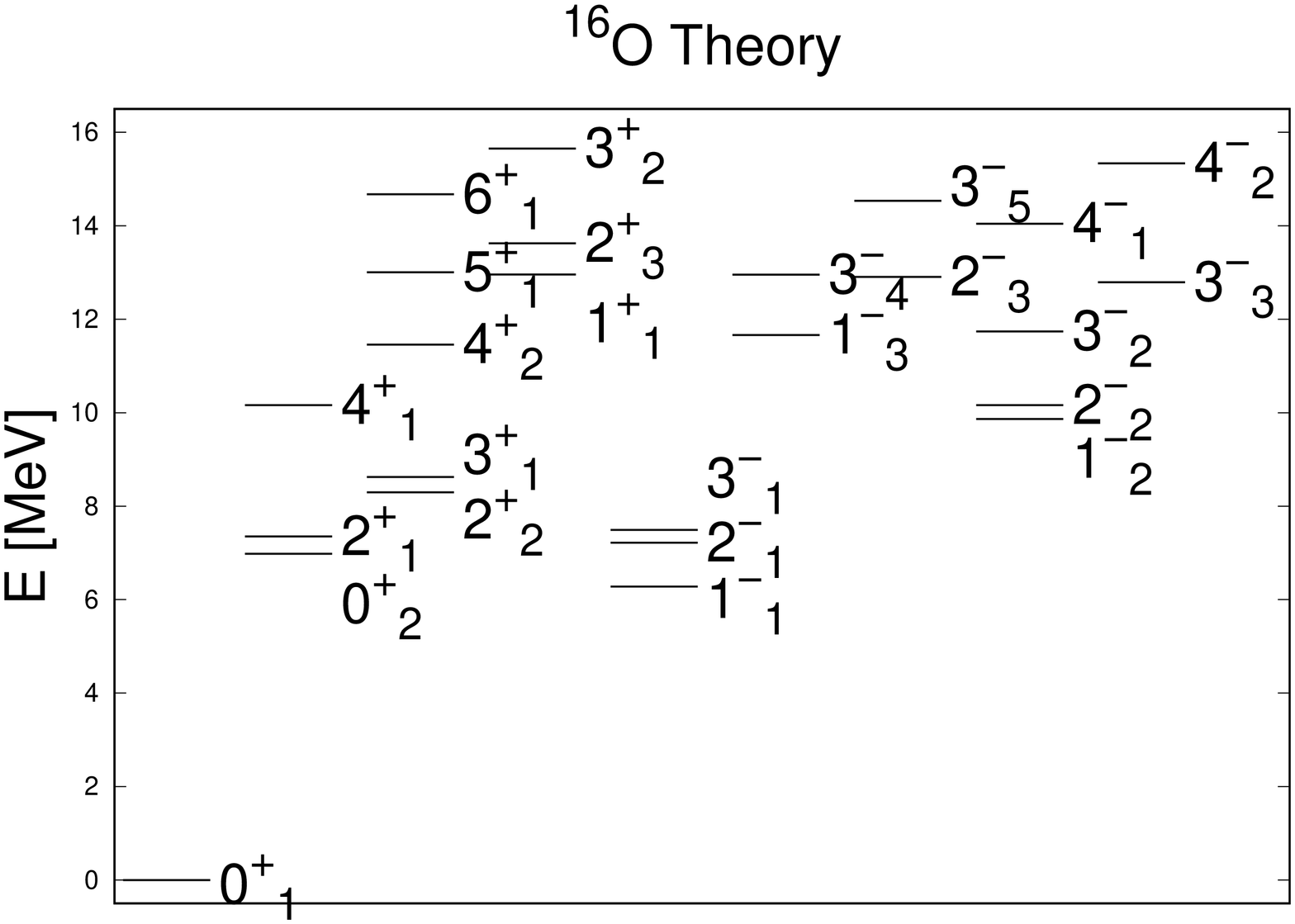}}} \\
\rotatebox{270}{\resizebox{120pt}{120pt}{\includegraphics[width=0.23\textwidth]{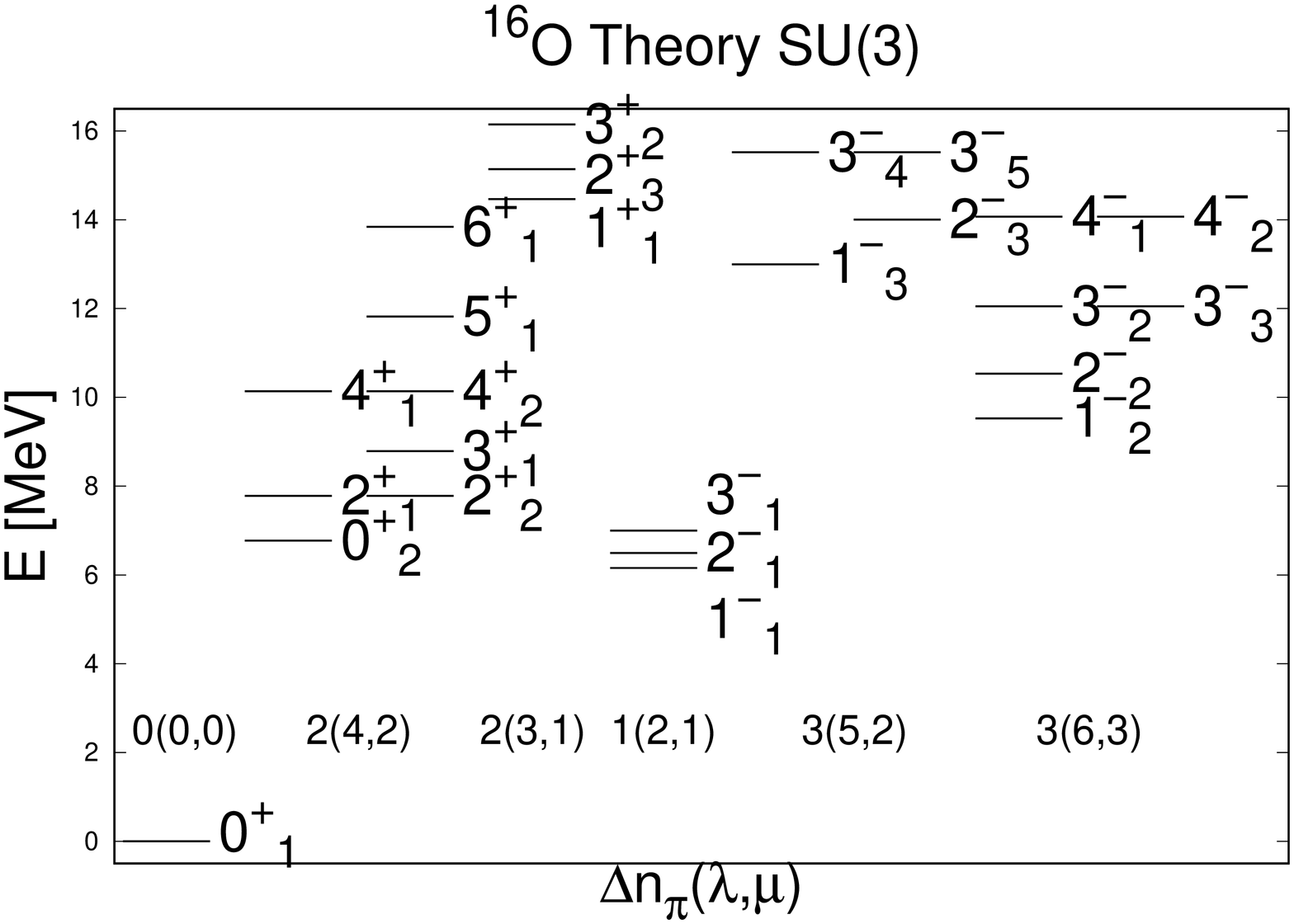}}}
\rotatebox{270}{\resizebox{120pt}{120pt}{\includegraphics[width=0.23\textwidth]{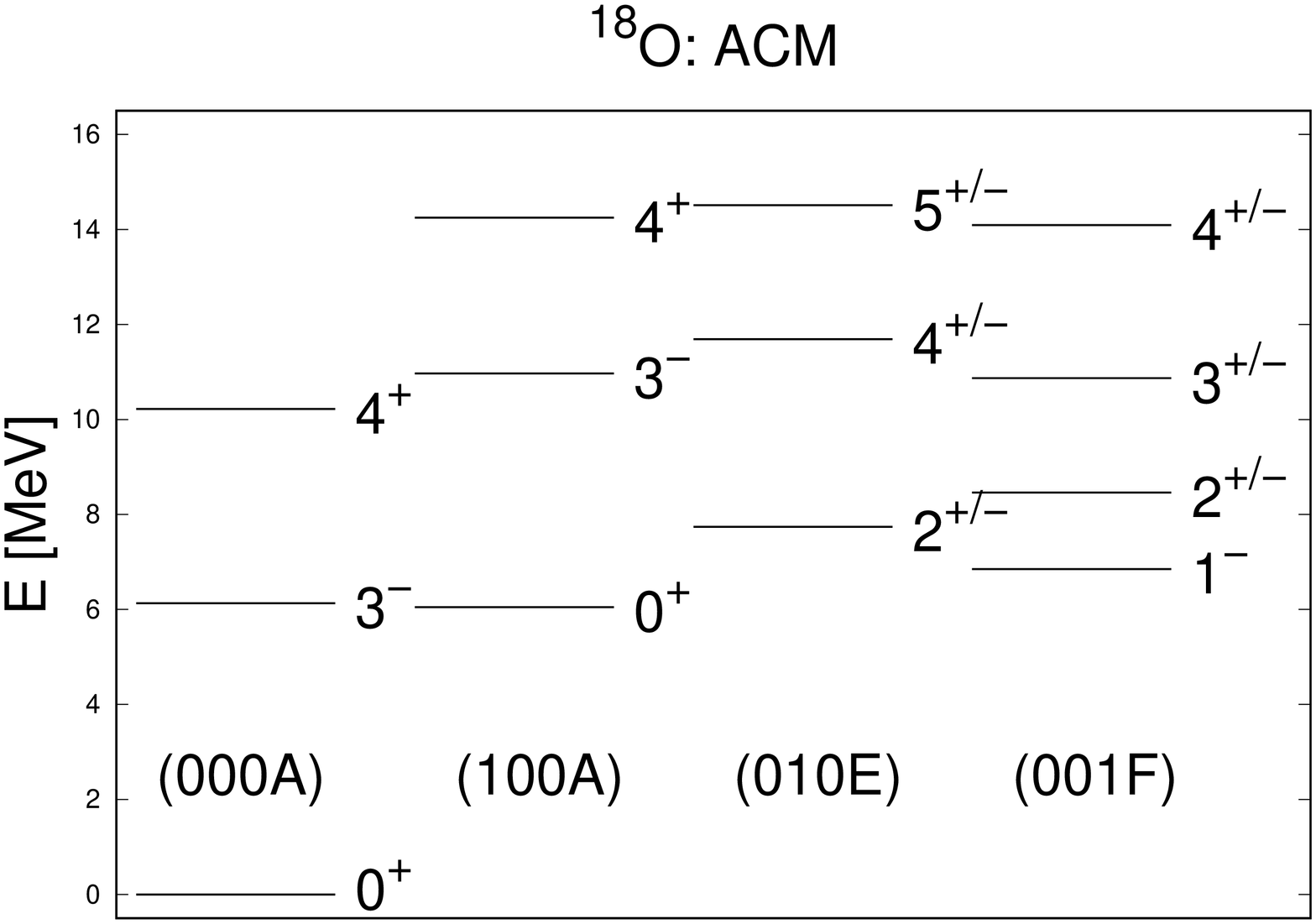}}} 
\caption{\label{fig1}
Spectrum of $^{16}$O. 
The left panel in the first row
 is the experimental spectrum and the right
one depicts the result of the theory. Experimental data are taken from \cite{exp}.
In the lower row, the left panel depicts the result for the $SU(3)$ limit while the right figure shows the result of
Ref. \cite{bijker2017}, with the ordering of bands as done in \cite{bijker2017}. The figure contains less bands than shown
in the other figures.
}
\end{center}
\end{figure}

In Fig. \ref{fig1}, on the left panel in the upper row
the experimental spectrum is depicted and the right hand figure in
the first row the theoretical spectrum, as obtained by our fit. In the lower row, left side, the result
for the $SU(3)$ limit is given and the last panel shows the result of \cite{bijker2017}.
There, the bands are ordered as done in \cite{bijker2017} and it contains less bands than listed in the other figures.

The agreement of the theoretical spectrum to the experimental one is satisfactory. The difference to the $SU(3)$
limit is minor, mainly the degeneracy between the two lowest $4^+$ states is lifted.
Comparing the $SU(3)$ limit with the calculation with mixing, the spectra look very similar. Inspecting Table \ref{tab2}
one also notes that the mixing term $b_1$ is very small. Another feature is that the $a$-parameter is zero, only
the $a_{Lnp}$-parameter remains and it is positive. This again illustrates that the $^{16}$O nucleus is spherical, with
the ground state band consisting only of the $0_1^+$ state. The exited states already belong to shell 
excitations ($\Delta n_\pi >0$), the nucleus gets deformed and can rotate ($a_{Lnp} \neq 0$).

The energy spectrum in \cite{bijker2017} is obtained with the formula

\beqa
\Delta E & = & \omega_1 v_1 +\omega_2 v_2 +\omega_1 v_2 + B_{[v]}L(L+1)
~~~,  
\label{specroelof}
\eeqa
where $v_k=0,1,...$ are the quantum numbers of the model and $B_{[v]}$ ($[v]$ = $v_1v_2,v_3$)
is the parameter to adjust for each band 
the moment of inertia. Four bands are considered: $(000A)$, $(100A)$, $(010E)$ and $(001F)$. The parameter values used are $\omega_k=6.05$MeV ($k=1,2,3$), $B_{000A}=0.511$MeV, $B_{100A}=0.410$MeV, $B_{010E}=0.282$MeV and
$B_{001F}=0.420$MeV. With this choice the spectrum depicted in Fig. \ref{fig1} on the right in the lower row is reproduced.
Note, that the formula (\ref{specroelof}) consists of {\it linear independent parameters}, which allow to adjust the states
within any band and, thus, is not of great predictive power.  On the other hand, in the SACM the parameters are interconnected
to all bands and thus are not linear independent, changing one parameter has an effect on all bands.

\begin{table}
\begin{center}
\begin{tabular}{|c|c|c|c|}
\hline
param. [MeV] & Theo. & $SU(3)$ \\
\hline
$\chi$ &  0.41263 & 0.46007 \\
$\xi$ &  0.26145 & 0.072041 \\
$t_1$ & -0.030068 & -0.019610  \\ 
$t_2$ & 0.0072660 & 0.00413550  \\ 
$a$ & 0. & 0.\\ 
$a_{Lnp}$ &  0.091400 & 0.0841543 \\ 
$b$ & -0.1998& 0.\\ 
$b_1$ & -0.051895 & 0. \\ 
$pe_2$ & 0.53112 & 0.33189 \\
\hline
 \end{tabular}
\end{center}
\caption{\label{tab2}
List of the parameter values used. The first column lists the parameter
symbols, the second their numerical value for the theoretical fit, including the $b_1$ term.
The third list are the parameters in the $SU(3)$ limit. Note the small $b_1$ parameter, which indicates a very small mixing to 
$SO(4)$.
}
\end{table}

\begin{table}
\begin{center}
\begin{tabular}{|c|c|c|c|c|c|}
\hline
$B(EL;J^\pi_i \rightarrow J^\pi_f)$ & EXP. & theo & $SU(3)$  & Ref. \cite{bijker2017} \\
\hline
$B(E2;2^+_1 \rightarrow 0^+_1)$ & $3.1\pm 0.1$ & 0.003 & 0.0  & 10.9 \\ 
$B(E2;2^+_1 \rightarrow 0^+_2)$ & $27.0\pm 3.0$ & 18.55 & 2.52 & 2.50 \\ 
$B(E2;4^+_1 \rightarrow 2^+_1)$ & $65.\pm6.0 $ & 0.06 & 0.62  & - \\ 
$B(E2;4^+_2 \rightarrow 2^+_1)$ & $1.\pm5.0 $ & 0.21 & 1.26  & 15.0 \\ 
$B(E2;1^-_1 \rightarrow 3^-_1)$ & $21.\pm5.0 $ & 0.& 0.0  & 7.93 \\ 
$B(E2;2^-_1 \rightarrow 1^-_1)$ & $10.3\pm 0.1$ & 11.16 & 4.36  & 3.34 \\ 
$B(E2;2^-_1 \rightarrow 3^-_1)$ & $8.2\pm 3.0$ & 11.56 & 4.52  & 4.18 \\ 
$B(E3;3^-_1 \rightarrow 0^+_1)$ & $13.5\pm 0.7$ & 20.08 & 21.45  & 5.27 \\ 
\hline
 \end{tabular}
\end{center}
\caption{\label{tab3}
List of $B(EL)$-transition values, measured and 
obtained in three different model calculations:
In the first column information is listed on the type of the electromagnetic
transition, the second column lists the corresponding experimental values,
the third column corresponds to the theoretical calculations with the $b_1$ term and the fourth column lists the 
values in the $SU(3)$ limit. The last column lists the values as obtained in \cite{bijker2017}.
}
\end{table}

In Table \ref{tab3} some $B(EL)$ transition values are listed and compared to experiment.
As can be seen, the $B(E2)$-values, which are a good measure for the structure of the wave functions, 
are well reproduced and often better than in the 
ACM \cite{bijker2017}. 
We use a geometrical estimation of the effective charges, as explained in \cite{huitzilin2012}. For the octupole transition
no additional fitting is used,
while for the B(E2)-transitions one global factor is adjusted. If this factor is near to the value of 1, it is
an indication that the geometrical estimation is quite good. As shown in \cite{huitzilin2012} the geometric estimate
of the effective charge is multiplied by a parameter $pe2$ ($\sim \beta^6$ \cite{bijker2017}) in the quadrupole operator. Consulting Table {\ref{tab2}, the $pe2$ is 
approximately 0.53, which indicates a rather good estimation.  
The $B(E3;3_1^- \rightarrow 0_1^+)$-value is
reproduced in order, but when the scaling factor $\sim \beta^8$ \cite{bijker2017} for the $B(E3)$ is applied, this value reduces to 8.63, which is 
in quite good agreement to experiment.
For the calculation
of this value an octupole deformation for the $3_1^-$ state was needed as an input. This value is estimated within 
the geometrical model \cite{greiner} and given in the Appendix. In order to have a better test of a model, one also
has to compare various $B(E3)$ and $B(E4)$ values to experiment and not just one. For our purposes this is not necessary, because it is sufficient to show if the PEP already leads in the spectrum  to clear structural differences to the ACM. 

In Tables \ref{tab4} and \ref{tab5} we list the content 
of some of the states in the spectrum in order to show that one can 
group them into bands. Shown is the percent contribution of a state to a given $n_\pi (\lambda , \mu )$: 
If a state belongs to the same band, the distribution has to be similar!
As can be seen, the $0_1^+$ state is the only state of a "band", it is a hundred percent pure (0,0).
These tables also show that all states are practically pure $SU(3)$ states, demonstrating that the $^{16}$O is a
show-case for the $SU(3)$ shell model. There are some small admixtures to other $SU(3)$ irreps, but the 
numbers listed are rounded off, such that e.g. 99.8 percent is shown as a 100 percent contribution.

One notes that the grouping into bands has nothing to do with the one defined in \cite{bijker2017}. In order to
associate states to the same band, the internal structure has to be the same (at least approximately with mixing). The
ground state belongs to the (0,0) $SU(3)$ irrep and 12 oscillation quanta, thus it is the {\it only}
representative of the band. The other bands are ordered according to the membership to $SU(3)$ irreps
and different irreps have distinct deformations. It is clear that
this ordering into bands does not agree at all with \cite{bijker2017}, where a pure classical picture of the nucleus
was assumed and the $\alpha$-particles with no internal structure. The association into bands
is maintained when mixing is included .

The experimental data can be reproduced better within the SACM (see Table \ref{tab3}). The spectrum can be
reproduced equally well to the already existing data. However, comparing the complete spectrum of both theories
leads to significant differences between the SACM and \cite{bijker2017}: Not considering the
PEP leads to a denser spectrum, which is lifted in the SACM. The multiplets reported 
in the (010E) and (001F) badns (see Fig. \ref{fig1} \cite{bijker2017} are not
there. Only the search for a more complete spectrum can show
the differences. The association into bands is easier in $^{16}$O than in $^{12}$C \cite{hess2018}, because the
states are each concentrated in practically one $SU(3)$ irrep.

\begin{table}
\begin{center}
\begin{tabular}{|c|c|c|c|c|c|c|}
\hline
n$\hbar\omega$ ($\lambda , \mu$) / $L^\pi_i$ & $0_1^+$ & $2_1^+$ & $4_1^+$ & $0_2^+$ & $2_2^+$ & $4_2^+$\\ 
\hline
$0$ :(0,0) & 100 & 0 & 0 & 0 & 0 & 0 \\
$2$ :(2,0) & 0 & 0 & 0 & 0 & 0 & 0 \\ 
$2$ :(3,1) & 0 & 0 & 0 & 0 & 0 & 1 \\ 
$2$ :(4,2) & 0 & 100 & 100 & 100 &100 & 99 \\ 
\hline
 \end{tabular}
\end{center}
\caption{\label{tab4}
$SU(3)$ content of some low lying states with positive parity, given
in percent, for the theoretical calculation which includes the mixing. 
The numbers are only approximate and not all irreps are shown. The $n\hbar\omega$
denotes the shell excitation number.
}
\end{table}

\begin{table}
\begin{center}
\begin{tabular}{|c|c|c|c|c|c|}
\hline
n$\hbar\omega$ ($\lambda , \mu$) / $L^\pi_i$ & $1_1^-$ & $2_1^-$ & $3_1^-$ \\ 
\hline
$1$ :(2,1) & 100 & 100 & 100  \\
$3$ :(3,0) & 0 & 0 & 0\\ 
$3$ :(4,1) & 0 & 0 & 0  \\ 
$3$ :(5,2) & 0 & 0 & 0o  \\ 
$3$ :(6,3) & 0 & 0 & 0  \\ 
\hline
 \end{tabular}
\end{center}
\caption{\label{tab5}
$SU(3)$ content of some low lying states with negative parity, given 
in percent, for the theoretical calculation which includes mixing.
The numbers are only approximate and not all irreps are shown.
The $n\hbar\omega$ denotes the shell excitation number.
}
\end{table}

\section{Conclusions}
\label{sec4}

The structure of $^{16}$O in terms of a 4-$\alpha$ particle system within the SACM and the importance of
the PEP was investigated, comparing it to \cite{bijker2017} where the PEP was ignored. It is shown that in the ground state 
a tetrahedral structure, after applying a geometric mapping, results as default using simple symmetry arguments. Ignoring the PEP leads to a denser spectrum at low energy, as already shown for the case of $^{12}$C in \cite{hess2018}. As a result, the degeneracy obtained in \cite{bijker2017} in some bands are not there at all. For example, the spin doublets in the 
(010E) and (001F) bands, predicted in the ACM, are broken and states are shifted to higher energy due to the
 implementation of the PEP.

The SACM can reproduce the experimental data well and for the transition values often better than in \cite{bijker2017},
even considering that it uses a phenomenological, algebraic Hamiltonian and not a microscopic interaction as used in 
\cite{DrO16,dytrych,THSR}. The algebraic structure makes it easier to compare to \cite{bijker2017} and also allows to retrieve 
more data.

In conclusion, one cannot ignore the PEP in $^{16}$O, otherwise it leads to a wrong understanding of the cluster structure.  
The SACM is able to describe well the structure of $^{16}$O, concerning the spectrum and several transition
values. The ACM, in its present form, is not applicable to the cluster structure of light nuclei.

Finally, though density profiles show an approximate tetrahedral structure \cite{kanada} and a classical treatment suggests a
grouping of bands according to the $T_4$ dynamical group, the antisymmetrization destroys it and requires a different
grouping into bands. Also states of a $T_4$ irrep are destroyed or shifted to higher energy.While the ACM works for
atomic cluster molecules, it cannot be applied in its present form to nuclear cluster systems, until the PEP is taken into account.

\section*{Acknowledgments}
We acknowledge financial support form DGAPA-PAPIIT  (IN100418) 
and CONACyT (project number 251817).

\section*{Appendix}

Here, we will construct a {\it geometrical} octupole oscillator model for $^{16}$O:

The Hamiltonian is given by

\beqa
{\bd H} & = & \hbar \Omega_3 \left( {\bd N} + \frac{7}{2}\right)
~~~,
\eeqa
where ${\bd N}$ is the number operator. The energy $\hbar\Omega_3$ of the first vibrational state $3_1^-$
can immediately be determined. The energy of this state is at 6.13~MeV \cite{exp}, i.e 

\beqa
\hbar\Omega_3 & = & 6.13~{\rm MeV}
~~~.
\eeqa

The geometrical quadrupole operator is to lowest order in $\alpha_{30}$  given by

\beqa
{\bd Q}_{30} & = & \frac{3Ze}{4\pi} R_0^3 {\bd \alpha}_{30}
~~~,
\eeqa
where $R_0=1,2~A^\frac{1}{3}$~fm, $e$ is the unit charge and ${\bd \alpha}_{30}$ is the deformation operator.

The $B(E3;3_1^- \rightarrow 0_1^+)$ value is given by

\beqa
B(E3;3_1^- \rightarrow 0_1^+) & = & \frac{1}{7}\mid \langle 0_1^+,0\mid\mid {\bd Q}_{30} \mid\mid 0_1~ +,0\rangle\mid^2
~~~,
\nonumber \\
\eeqa
where the "1"=$(2J_f+1)$, with $J_f=0$ and "7"=$(2J_i+1)$, with $J_i=3$, the final and initial spin of the states. 

For the calculation of the reduced matrix element, we define first the initial and final states, furthermore the quantization
of the variable
${\bd \alpha}_{30}$:

\beqa
\mid i \rangle & = & {\bd b}^\dagger_{30} \mid 0 \rangle
\nonumber \\
\mid f \rangle & = & \mid 0 \rangle
\nonumber \\
{\bd \alpha}_{30} & = & \sqrt{\frac{\hbar}{2B_3\Omega_3}} \left( {\bd b}^\dagger_{30} +{\bd b}_{30}\right)
~~~,
\eeqa 
with ${\bd b}^{30}={\bd b}_{30}$ and $B_3$ is the mass parameter of the geometric theory.

With this, the matrix element of the quadrupole operator is

\beqa
\langle f \mid {\bd Q}_{30} \mid i \rangle & = & \frac{3Ze}{4\pi} R_0^3 \sqrt{\frac{\hbar}{2B_3\Omega_3}}
\nonumber \\
& = &
\left( 
\begin{array}{ccc}
0 & 3 & 0 \\
0 & 0 & 0
\end{array}
\right)
\langle f\mid\mid {\bd Q}_3\mid\mid i\rangle 
\nonumber \\
& = & -\frac{1}{\sqrt{7}}\langle f\mid\mid {\bd Q}_3\mid\mid i\rangle
\eeqa
($R_0=1.2~A^{\frac{1}{3}}$). Therefore, the reduced matrix element is

\beqa
\langle f\mid\mid {\bd Q}_3\mid\mid i\rangle & = & \frac{3Ze}{4\pi} R_0^3 \sqrt{\frac{\hbar}{2B_3\Omega_3}}
\left(-\sqrt{7}\right)
~~~.
\eeqa
Using the expresion for the $B(E3)$ value above and resolving for $B_3c^2$ (using $\frac{\hbar}{B_3\Omega_3}$ =
$\frac{(\hbar c)^2}{(B_3c^2)(\hbar\Omega_3)}$) , we obtain

\beqa
&
-\sqrt{7}\left(\frac{3 Z e}{4\pi}\right) R_0^3 \sqrt{\frac{(\hbar c)^2}{2(\hbar \Omega_3)}}\frac{1}{\sqrt{B_3c^2}}
& 
\nonumber \\
&
= 13.5~(0.05940~A^{\frac{4}{3}})^{\frac{1}{2}}~{\rm e}~{\rm fm}^3~=20.89~{\rm e}~{\rm fm}^3
&
~~~,
\eeqa
where we use the experimental value 13.5~WU for the octupole transition on the right hand side and the factor in the
parenthesis is the conversion from WU to ${\rm e}^2~{\rm fm}^4$.

Plugging in the values and resolve for $(Bc^2)$, also using $e^2=1.44$MeVfm, we obtain the value

\beqa
\left(B_3c^2\right) & \approx & 204555~{\rm MeV}{\rm fm}^2
~~~.
\eeqa

Now we determine the value of $\alpha_{30}$, which gives the octuplole deformation denoted by $\beta_{30}$, 
defined as

\beqa
\beta_{30} ~ = ~ \alpha_{30} & = & \langle f \mid {\bd \alpha}_{30}\mid i \rangle ~ = ~
 \sqrt{\frac{(\hbar c)^2}{2(B_3c^2)(\hbar \Omega_3)}}
\nonumber  \\
& = & 0.125
~~~,
\eeqa
which gives the same as $\langle f \mid ({\bd \alpha}_{30})^2\mid i \rangle$.


\begin{thebibliography}{99}

\bibitem{DrO16} K. D. Launey, T. Dytrych and J. P. Draayer, Progr. in Part. and Nucl. Phys. {\bf 89} (2016), 101.

\bibitem{dytrych} T. Dytrych, {\it Evidence for Symplectic Symmetry in ab initio no-core Shell Model Results}, (PhD Thesis,
Lousiana State Universiuty, 2008).

\bibitem{schuck2013} P Schuck, J. Phys.: Conf. Ser. {\bf 436} (2013),  012065

\bibitem{schuck2017} P. Schuck, Y. Funaki, H. Horiuchi, G. R\"opke, A. Tohsaki and T. Yamada, 
Physica Scripta {\bf 91} (2016), 123001.

\bibitem{funaki2018} Y. Funaki, Phys. Rev. C {\bf 97} (2018), 021304(R).

\bibitem{THSR} A. Tohsaki, H. Horiuchi, P. Schuck and G. R\"opke,
Phys. Rev. Lett. {\bf 87} (2001), 192501.

\bibitem{freer2017}M.  Freer, Rep. Prog. Phys. {\bf 70} (2007) 2149.

\bibitem{review-clusters} M. Freer, H. Horiuchi, 
Y Kanada-En'yo, D. Lee and U.-G. Meissner, Rev. Mod. Phys. {\bf 90} (2018), 03500.

\bibitem{funaki2010} Y. Funaki, M Girod, H Horiuchi, G R\"opke, P Schuck,
A Tohsaki and T Yamada, J. Phys. G: Nucl. Part. Phys. {\bf 37} (2010) 064012.

\bibitem{schuck2018} P. Schuck, AIP Conference Proceedings {\bf 2038} (2018), 020002

\bibitem{bijker2016} R. Bijker and F. Iachello, Phus. Rev, Lett. {\bf 112} (2014), 152501.

\bibitem{bijker2017} R. Bijker and F. Iachello, Nucl. Phys. A {\bf 957} (2017), 154.

\bibitem{bijker2014} D. J. Mar\ii n-L\'ambarri, R. Bijker, M. Freer, et al., Phys. Rev. Lett.
{\bf 113} (2014), 012502.

\bibitem{annphys} R. Bijker and F. Iachello, Ann. Phys. (N.Y.) {\bf 298} (2002), 334.

\bibitem{hess2018} P. O. Hess, Eur. Phys. J. A {\bf 54} (2018), 32.

\bibitem{sacm1} J. Cseh, Phys. Lett. B \textbf{281} (1992), 173.

\bibitem{sacm2} J. Cseh and G. L\'evai, Ann. Phys. (N.Y.) \textbf{230}
(1994), 165.

\bibitem{hoyle} F. Hoyle, Ap. J. Suppl. {\bf 1} (1954), 121.

\bibitem{wildermuth} K. Wildermuth and Y. C. Tang, {\it A Unified Theory of the
Nucleus}, (Academic Press, New York, 1977).

\bibitem{kramer-II} V. C. Aguilera-Navarro, M. Moshinsky and P. Kramer, Ann. Phys. {\bf 54} (1969), 379.

\bibitem{mosh-book} M. Moshinsky and Y. Smirnoc. {\it The Harmonic Oscillator inModern  Physics}, (Harwood Academic Publishers,
Australia, 1996).

\bibitem{supermultiplets}E. Wigner,  in {\it Group Theoretical Concepts and Methods in Elementary Particle Physics},
ed. by. F. G\"ursey.,  (Gordon and Breach, New York, 1964).

\bibitem{aka} J.P. Draayer and Y. Akiyama, J. Math. Phys. {\bf 14} (1973), 1904.

\bibitem{bahri} D. J. Rowe and C. Bahri, J. Math. Phys. {\ bf 41} (2000), 6544.

\bibitem{elliott} J. P. Elliott, Proc. R. Soc. London A {\bf 245}, (1958) 128; 
{\bf 245} (1958), 562.

\bibitem{rowe} D J Rowe, Rep. Prog. Phys. {\bf 48} (1985), 1419.

\bibitem{castanos1988} O. Casta\~nos, J. P. Draayer and Y. Leschber, Z. f. Phys. A{\bf 329}  (1988), 33.

\bibitem{hw} J. Blomqvist and A. Molinari, Nucl. Phys. A {\bf 106} (1968), 545.

\bibitem{octavio} O. Casta\~nos and J. P. Draayer, Nucl. Phys. A {\bf 491} (1989), 349.

\bibitem{sympl} O. Casta\~nos, P. O. Hess, P. Rocheford and J. P. Draayer, Nucl. Phys. A {\bf 524} (1991), 469.

\bibitem{geom} P. O. Hess, G. L\'evai and J. Cseh, Phys. Rev. C {\bf 54} (1996), 2345.

\bibitem{kanada} Y. Kanada-En'yo, Phys. Rev. C {\bf 96} (2017), 034306.

\bibitem{exp} www.nndc.bnl.gov/ensdf

\bibitem{huitzilin2012} H. Y\'epez-Mart\ii ınez, M. J. Ermamatov, P. R. Fraser and P. O. Hess, Phys. Rev. C 
{\bf 86} (2012), 034309.

\bibitem{greiner} W. Greiner and J. M. Eisenberg, {\it Nuclear Theory I: Nuclear Models},
(North-Holland, Amsterdam, 1987).

\end{thebibliography}
\end{document}